\documentclass[12pt]{amsart}
\usepackage{amsmath}
\usepackage{amssymb}
\usepackage{mathrsfs}
\usepackage{graphicx}
\usepackage{url}
\usepackage{hyperref}
\usepackage{color}
\definecolor{refcolor}{RGB}{0,0,190}
\hypersetup{
    colorlinks,
    citecolor=refcolor,
    filecolor=refcolor,
    linkcolor=refcolor,
    urlcolor=refcolor
}

\newcommand{\schrod}{Schr\"o\-ding\-er}

\newcommand{\ket}[1]{|#1\rangle}

\newcommand{\qmU}{$\mathscr{U}$}
\newcommand{\qmR}{$\mathscr{R}$}

\def\({\left(}
\def\){\right)}

\newcommand{\eg}{\textit{e.g.} }

\newcommand{\imagew}[3]{
\begin{figure}[!ht]
\begin{center}
\includegraphics[width=#2\textwidth]{img/#1}
\caption{\small{\label{#1}#3}}
\end{center}
\end{figure}
}

\newtheorem{hypothesis}{Hypothesis}
\newtheorem{property}{Property}

\begin{document}
 
\title{Global and local aspects of causality in quantum mechanics
{\\
\vspace{0.1in}
\tiny
Talk delivered at the conference TM2012 -- ``The Time Machine Factory''}}
\author{Cristinel Stoica}
\date{October 16 2012, Turin}
\thanks{
\textit{Author email:} \href{holotronix@gmail.com}{holotronix@gmail.com}
\\
\textit{Conference web page:} \href{http://timemachine.polito.it/home/?q=node/2}{http://timemachine.polito.it/home/?q=node/2}
\\
\textit{Proceedings:} \href{http://www.epj-conferences.org/articles/epjconf/abs/2013/19/contents/contents.html}{http://www.epj-conferences.org/articles/epjconf/abs/2013/19/contents/contents.html}
\\
\textit{Abstract:} \href{http://www.epj-conferences.org/articles/epjconf/abs/2013/19/epjconf_tm2012_01017/epjconf_tm2012_01017.html}{http://www.epj-conferences.org/articles/epjconf/abs/2013/19/epjconf\_tm2012\_01017/\\epjconf\_tm2012\_01017.html}
}

\begin{abstract}
Quantum mechanics forces us to reconsider certain aspects of classical causality. The `central mystery' of quantum mechanics manifests in different ways, depending on the interpretation. This mystery can be formulated as the possibility of selecting part of the initial conditions of the Universe `retroactively'. This talk aims to show that there is a global, timeless, `bird's view' of the spacetime, which makes this mystery more reasonable. We will review some well-known quantum effects from the perspective of global consistency.
\end{abstract}
 
\maketitle
\tableofcontents

\section{Introduction}
\label{intro}

Quantum mechanics is well resumed by the \textit{unitary evolution process} {\qmU}, and by the \textit{state vector reduction process} {\qmR} (including \textit{the Born rule}, which gives the probabilities of the outcomes). This is correct, and free of paradoxes, when limiting to quantum mechanics from the Hilbert space perspective. 

When trying to account for reality, locality, or Lorentz invariance, strange correlations between apparently disconnected events appear.
That's why most interpretations introduce additional elements in discussion: hidden variables and instantaneous communications \cite{db1956double_solution_brief,Bohm52}, transactions which change the past \cite{cramer1988overview}, wavefunctions evolving back in time \cite{aharonov1964time,aharonov2007TSV,aharonov2012future-past}, splitting worlds \cite{Eve57} etc.

The {\qmR} process seems to be ad-hoc, apparently without a cause, violates the local conservation laws, and there is no decisive evidence supporting it \cite{schlosshauer2006experimental}. Is it really needed?

There are powerful reasons supporting the view that the {\qmR} process takes place discontinuously. If we try to explain this away in terms of unitary evolution {\qmU} we encounter serious difficulties. But it is important to see at least how much we can explain in terms of unitary evolution only, without appealing to discontinuous collapse.

This justifies the study of the following hypothesis:
\begin{hypothesis}[Unitarity hypothesis]
\label{hypo}
The {\qmR} process is reducible to the {\qmU} process.
\end{hypothesis}

Tegmark and Wheeler \cite{tegmark100quantum} mention a poll made at a conference on quantum computation, at the Isaac Newton Institute in Cambridge, in July 1999.
One of the questions was \textit{``Do you believe that all isolated systems obey the {\schrod} equation (evolve unitarily)?''}. 59 physicists answered ``yes'', 6 ``no'', and 31 were undecided.

This result is mainly due to the success of approaches which push the limit where the collapse should take place.
I would name here the Many Worlds Interpretation (MWI) \cite{Eve57}, the \textit{consistent histories} interpretation \cite{Gri84,Omn92}, and especially the \textit{decoherence program} \cite{joos-zeh1985emergence,Zur03a,Zur05,sch05}.

Other arguments for the unitarity hypothesis \ref{hypo} are presented by the author \cite{Sto08b,Sto08f,Sto12QMa}, and by 't Hooft \cite{hooft2011wave}. 

The main purpose of this talk is to advocate the idea that many quantum phenomena can be understood in terms of the unitary evolution {\qmU}, if we consider an important ingredient, the \textit{global consistency}.

Global consistency is not an additional element, it is intrinsic to the theory, but it did not receive enough attention in its explicit form. It is responsible for phenomena which appear to contradict our intuition about causality. In addition, it reduces the gap between the quantum mechanical view of a time evolution taking place in the Hilbert space, and the \textit{block world} view introduced by the theory of relativity.

\section{The global consistency principle}


Physical laws, including unitary evolution, are usually described by partial differential equations (PDE). A solution at $t_0$ can be extended into the future, but also into the past (fig. \ref{pde_initial_conditions_all.jpg} \textbf{A, B}). Hence, causality is bidirectional.
Special relativity shows that there is even more freedom, in choosing the hypersurface on which initial data is defined (fig. \ref{pde_initial_conditions_all.jpg} C).

\imagew{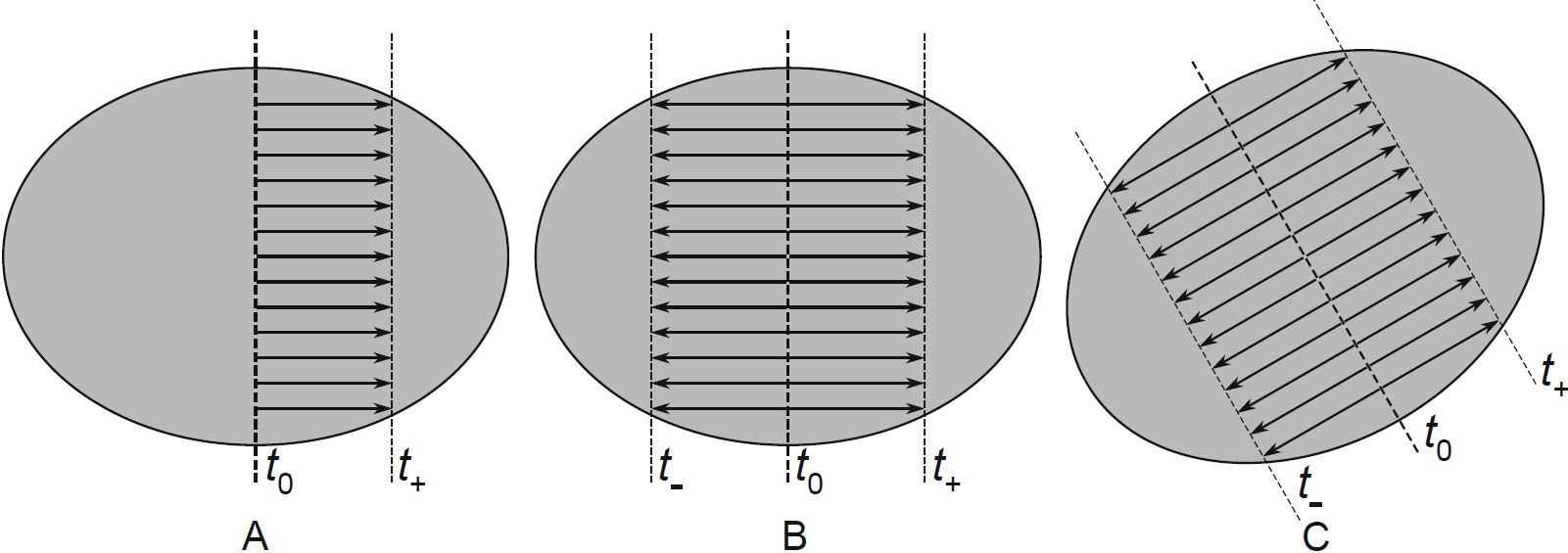}{0.675}{Initial conditions determine \textbf{A} the future, and \textbf{B} the past. \textbf{C.} The freedom to choose the initial surface.}

The \textit{local aspects of causality} refer to the way a solution extends (or propagates) in its neighborhood. 
Since causality is not always clearly defined, perhaps it is more appropriate to discuss about \textit{local consistency} instead.


The mathematical theory which deals with the passage from local to global solutions is named \textit{Sheaf Theory} \cite{bredon1997sheaf}.

A \textit{sheaf} is a rule to associate to an open set from a topological space a collection of objects named \textit{sections} -- \eg local solutions of a PDE, defined on that open set (fig. \ref{sheaves.jpg} \textbf{A}). By \textit{restricting} a section to a subset, we obtain a section. If two sections are equal on the intersection of their domains, they can be extended to the \textit{union} of their domains (fig. \ref{sheaves.jpg} \textbf{B}). Eventually, there may be a \textit{maximal domain} to which a section can be extended (fig. \ref{sheaves.jpg} \textbf{C}). 

\imagew{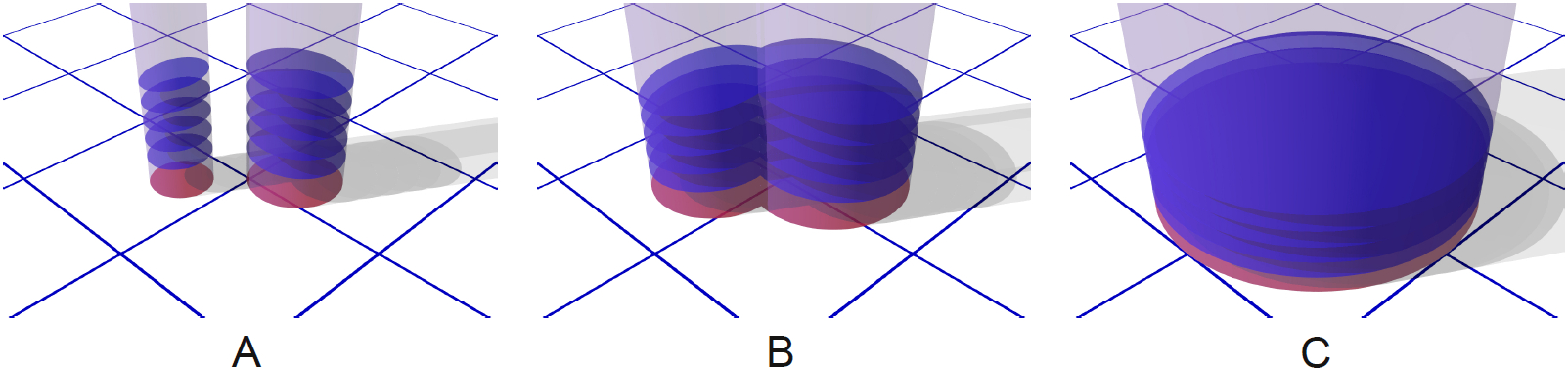}{0.8}{\textbf{A.} A sheaf associates to an open set a collection of local sections. \textbf{B.} Sections equal on the intersection of their domains can be glued on the union of their domains. \textbf{C.} There is a maximal domain to which a section can extend.}

But only few local solutions admit global extensions.
This means that \textit{global constraints} have something to say!


As an example, the wave equation imposes to the solution \textit{local constraints}.
A standing wave in a cavity has, in addition, \textit{global constraints}, in the form of boundary conditions (fig. \ref{standing_waves.jpg}).

\imagew{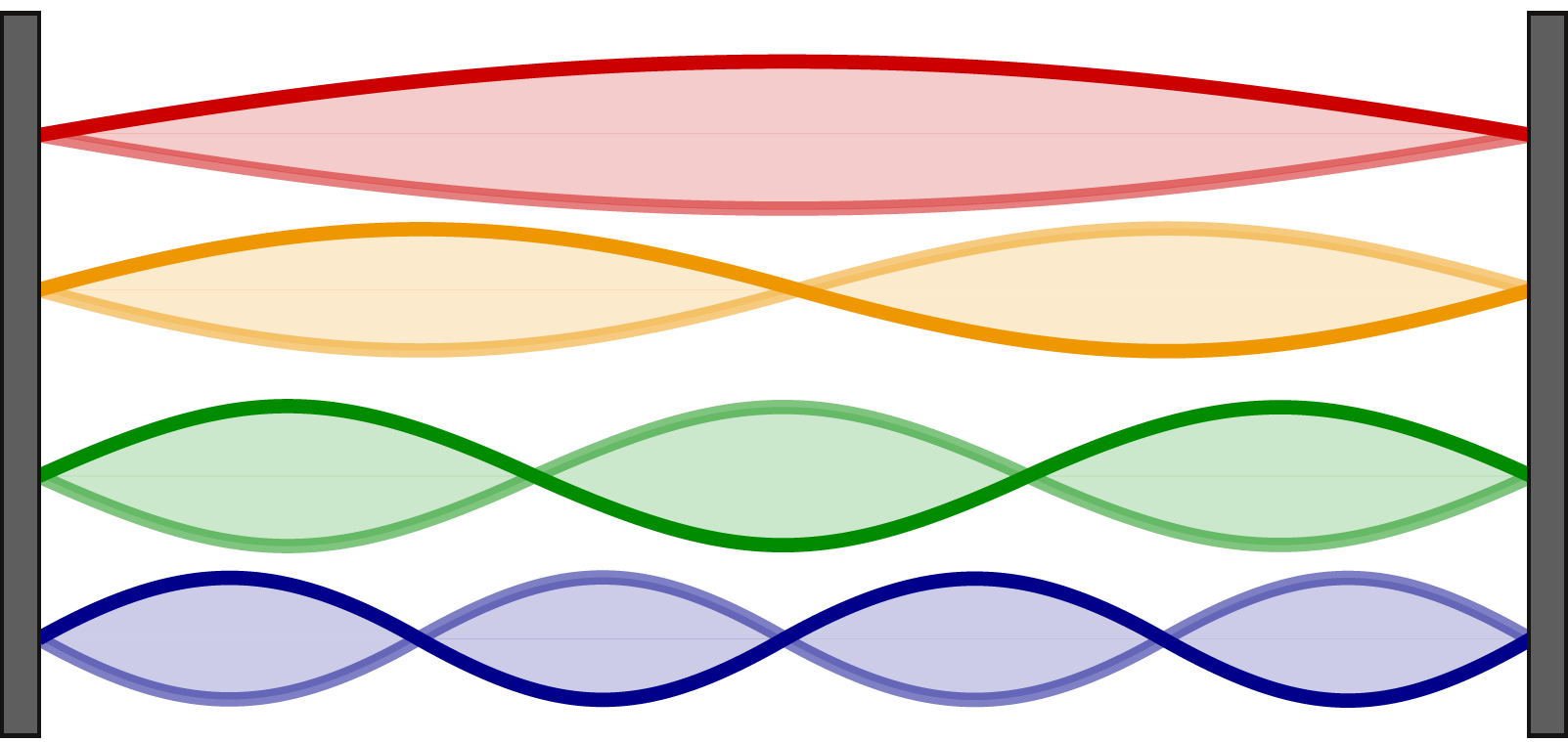}{0.4}{Standing waves in a cavity, an example of how only some local solutions can extend to global solutions.}

Louis de Broglie used this idea to explain the energy levels of the electron in the atom. He associated to the electron a wave, and required it to be a self-consistent standing wave \cite{dB24}. 

Inspired by de Broglie's idea, {\schrod} wrote the equation for the wavefunction -- the \textit{local constraint}. 
Then, he explained the discrete energy spectrum of the electron in the atom by \textit{global constraints}, in the form of boundary conditions on the sphere at infinity \cite{Sch26}.

From the viewpoint of \textit{initial conditions}, this may look strange. Initial conditions seem to ``know'' how they are allowed to be, to satisfy the boundary conditions at infinity. The conditions at different points are consistent, so that they can merge and extend to a global solution, which in turn satisfies boundary conditions at infinity!

This suggests the following order of precedence:
\begin{enumerate}
	\item
	The \textit{local consistency conditions} (\eg the PDE).
	\item
	The \textit{global consistency conditions} (\eg boundary conditions). From the possible solutions allowed by the local consistency condition, are selected only those satisfying also the global conditions.
	\item
	The \textit{initial conditions}. They have to be compatible with the local and global consistency conditions.
\end{enumerate}

To summarize, the local properties constrain the global properties, which in turn constrain the initial conditions.


The examples of the standing waves, and of the time independent {\schrod} equation, as local consistency conditions, and boundary conditions as global consistency conditions, apply to a wavefunction defined on space.
But similar global consistency take place as well for time-dependent fields defined on space+time.

In space+time, we should expect correlations between causally separated events, and between past and future.
Could the non-local behavior of quantum mechanics be better understood in terms of global consistency?

\section{Delayed initial conditions}

A quantum system evolves unitarily according to {\schrod}'s equation. The solution has the form
\begin{equation}
	\ket{\psi(t)}=U(t,t_0)\ket{\psi(t_0)},
\end{equation}
with initial condition $\ket{\psi(t_0)}=\ket{\psi_0}$.


The state of the system at a time $t_0$ is determined by a measurement. But we can't measure the state as it is, only an \textit{observable}. The outcome is an eigenvalue of the observable. The eigenvalue of the observable obtained when measuring it determines which \textit{eigenstate} -- that is, which of the possible solutions -- describes the system (fig. \ref{qm_outcomes_2_init_cond.jpg}).

\imagew{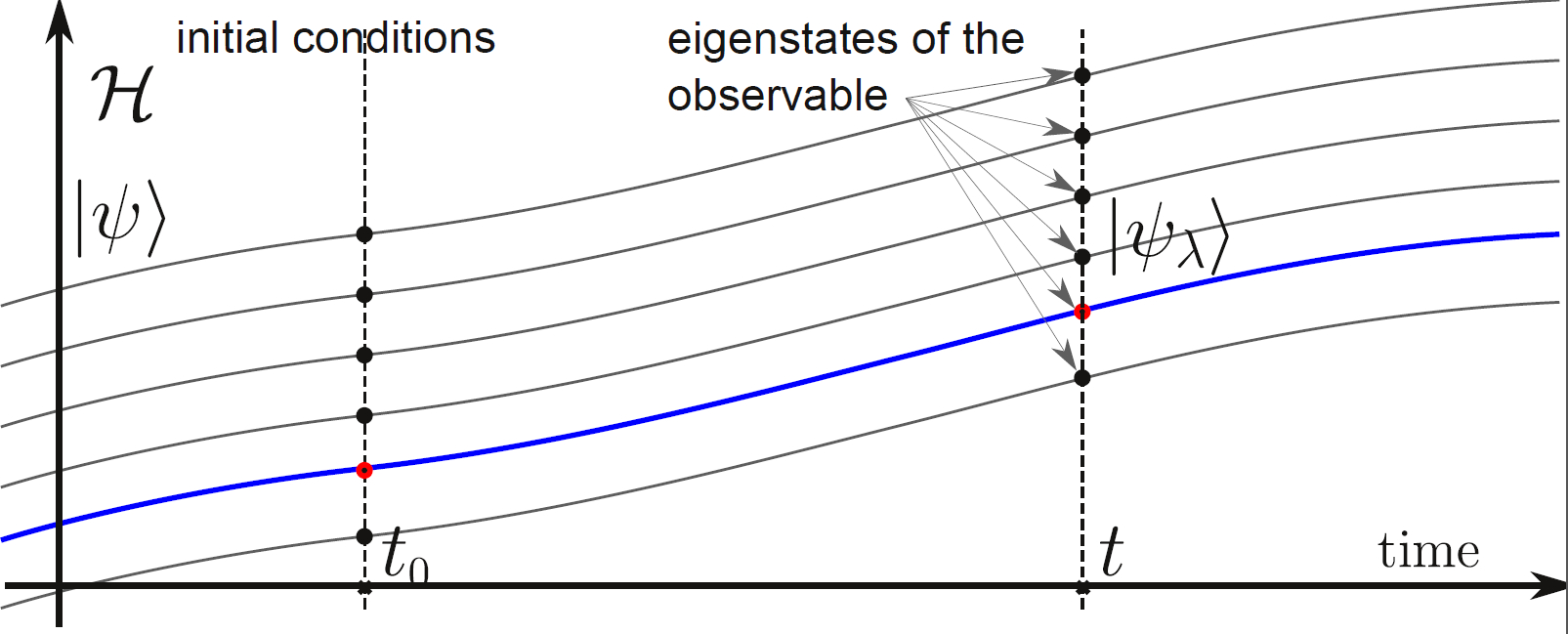}{0.65}{The initial states are constrained to evolve into eigenstates of the observable.}

For example, the spin of the electron in a magnetic field can only be $\ket{\uparrow}$ or $\ket{\downarrow}$ along the measurement direction, as in fig. \ref{sg_spin.png}. It cannot have other orientations, even though they are superpositions of $\ket{\uparrow}$ and $\ket{\downarrow}$.

\imagew{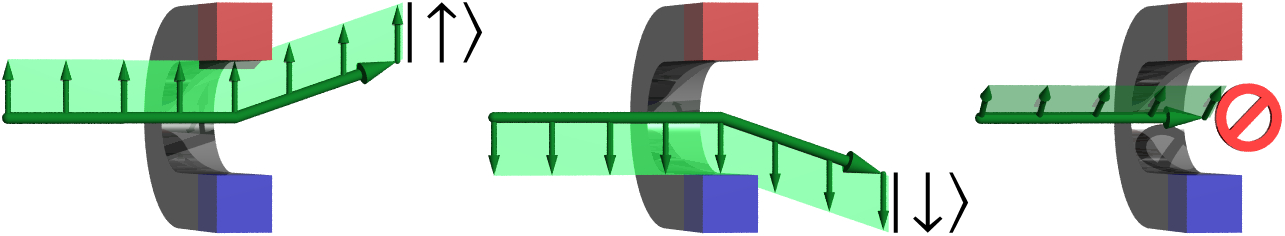}{0.75}{The spin can be $\ket{\uparrow}$, or $\ket{\downarrow}$, but not $\alpha\ket{\uparrow}+\beta\ket{\downarrow}$.}

Another observable may impose constraints which are \textit{incompatible} with those of the original observable. 
This suggests the following property of measurements \cite{Sto12QMb}:

\begin{property}[of restricted initial conditions]
\label{property_ic}
Not all possible initial conditions of the observed system can lead to definite outcomes of the measurement.
\end{property}

This property is proven mathematically to hold, under the unitarity hypothesis \ref{hypo}, even if we consider that the observation disturbes the system, and even by taking the environment into account \cite{Sto12QMb}. It also holds for hidden variable theories, as Bell's \cite{Bel64} and Kochen-Specker's \cite{kochen1967problem} theorems show. The choice of the observable selects the possible initial conditions, indefinitely in the past.

But doesn't this delayed selection of initial conditions violate causality? Not necessarily: each measurement reduces the set of the allowed solutions of {\schrod}'s equation, by eliminating the solutions incompatible with the observable, and \textit{not by changing the past} (fig. \ref{determinism-delayed.jpg}).
We call this phenomenon \textit{delayed initial conditions} \cite{Sto08b,Sto08f,Sto12QMa}.

\imagew{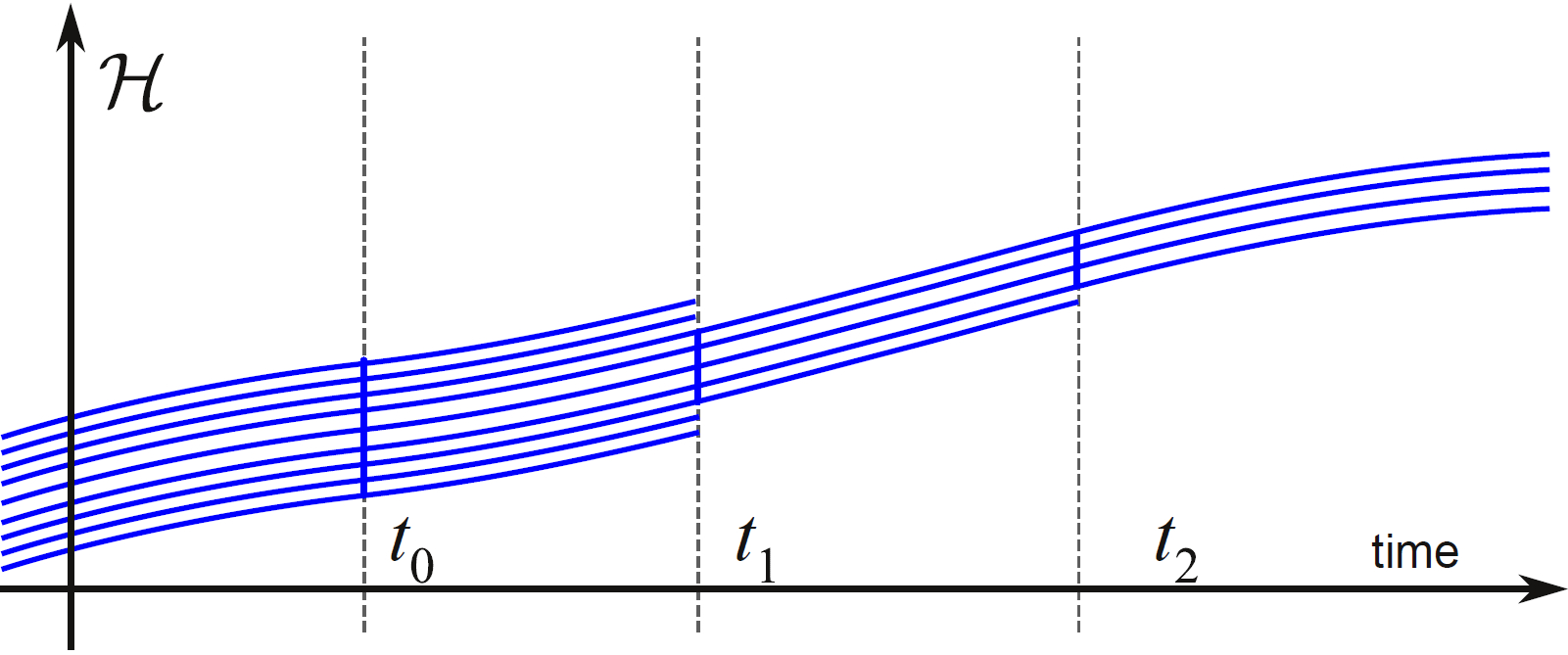}{0.65}{Delayed initial conditions.}

One may think that if we admit a discontinuous, or at least non-unitary {\qmR} process, the initial conditions are reset by the measurement, and property \ref{property_ic} no longer holds, or more precisely, it holds only since the last collapse.

But Wheeler's \textit{delayed choice experiments} show that the collapse can be pushed indefinitely back in the past \cite{Whe78,Whe83,delayed1986}. Suppose a laser ray is split by the first beam splitter of the Mach-Zehnder interferometer. The second beam splitter recomposes the ray, and all photons arrive at the detector \textbf{B} (fig \ref{mz_both_ways.jpg}).

\imagew{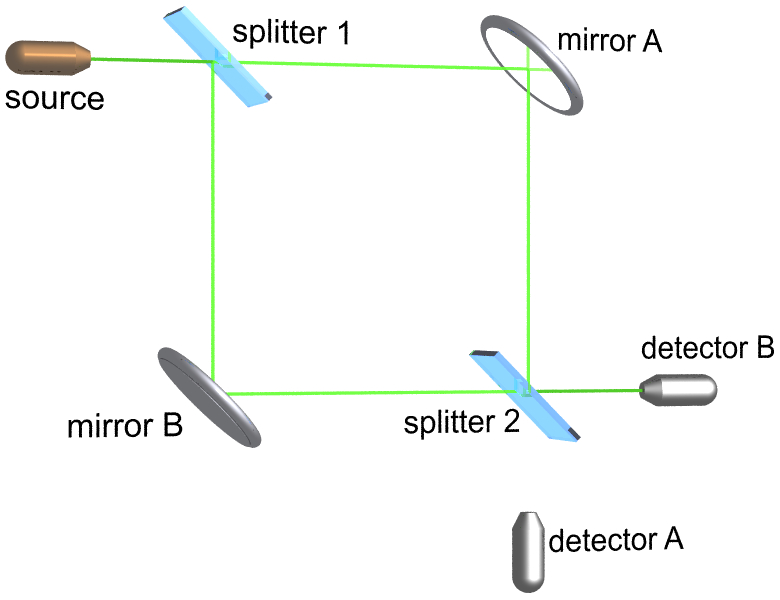}{0.55}{Both-ways observation.}
\imagew{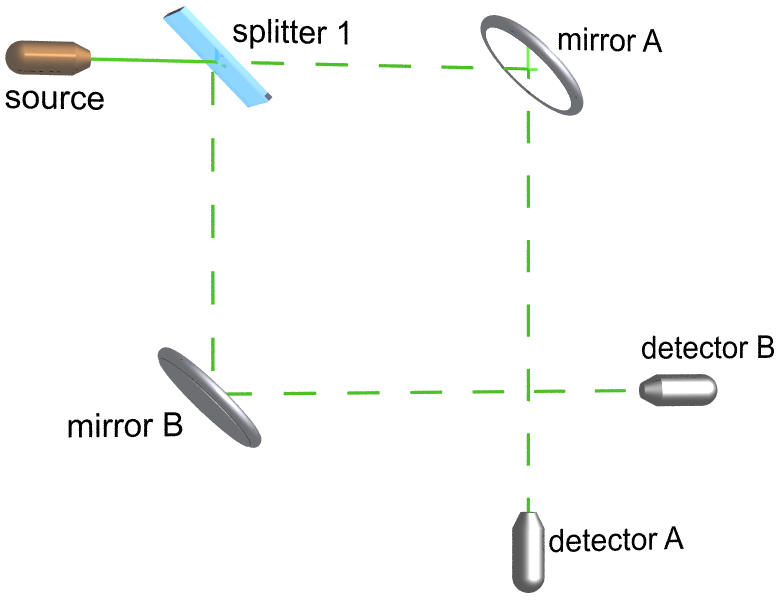}{0.55}{Which-way observation.}

If we remove the second beam splitter, the ray is no longer recomposed. Some photons will arrive at detector \textbf{A}, and others at detector \textbf{B} (fig \ref{mz_which_way.jpg}).

Our choice to remove beam splitter $2$ affects the way the ray is decomposed by splitter $1$,
even if we make the decision after the ray passed through the first splitter.

This shows that Property \ref{property_ic} cannot be completely avoided, even if we assume discontinuous collapse.

\section{Incompatible measurements}

Suppose that the spin is measured two times, along different directions in space. The two conditions are incompatible, and require a collapse (fig \ref{sg_spin_rot_collapse.jpg}).

\imagew{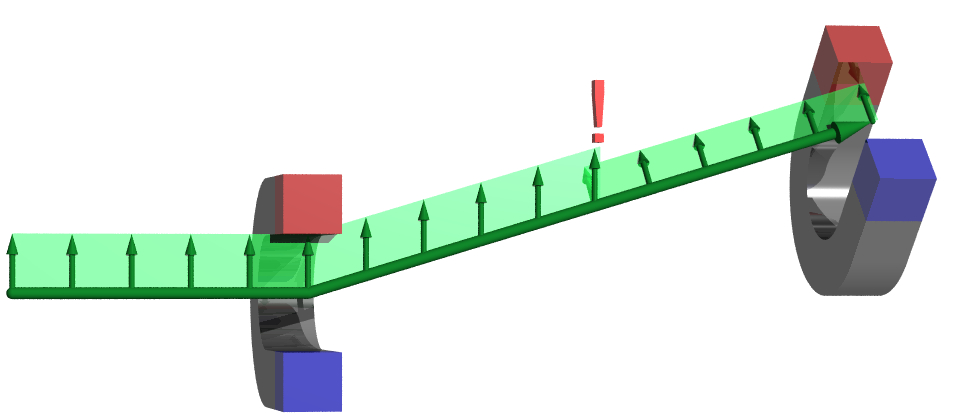}{0.65}{Incompatible measurements seem to require a discontinuous collapse.}

But it is not excluded that the interaction with the measurement device (and maybe with the environment) disturbed the observed system exactly as needed to obtain a correct outcome, without violating unitarity (fig. \ref{sg_spin_rot.jpg}). 

\imagew{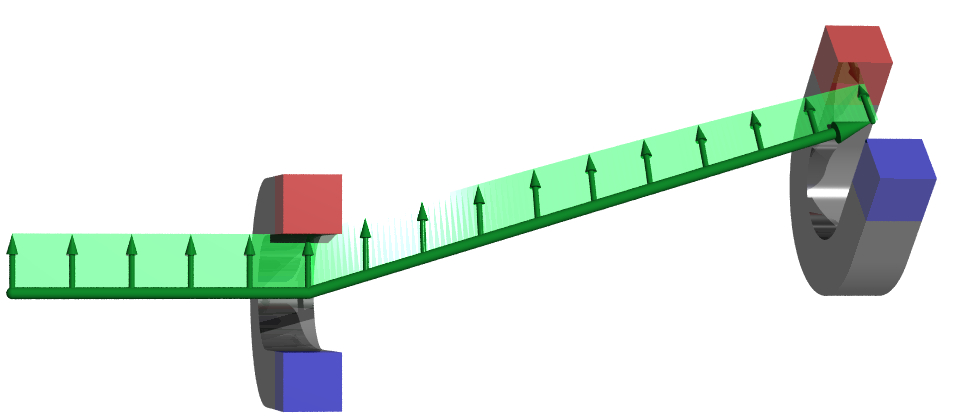}{0.65}{A unitary interaction may explain both results.}

Property \ref{property_ic}, proven in \cite{Sto12QMb}, shows that the disturbance caused by the measurement device cannot explain the outcome for any initial state. But we can still hope that the first measurement apparatus also interacted with the observed system, after the measurement, and the first and second disturbances combined may explain the result. 
The idea that the previous interaction disturbed the system in a way which depends on the future measurement is very strange from the viewpoint of causality.

What if the measurement is non-disturbing, and incompatible with the previous measurement (fig. \ref{qm_preduction.jpg})? How could the previous state evolve into the new state?

\imagew{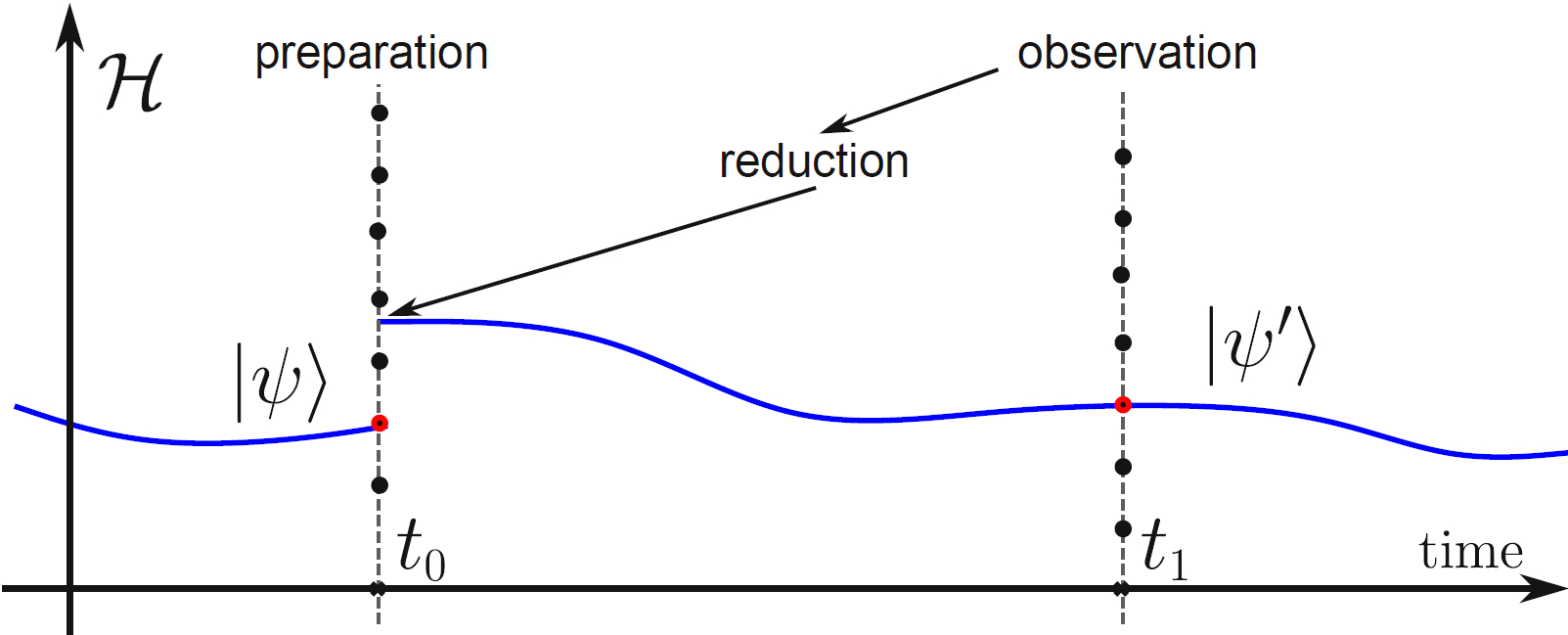}{0.65}{The apparent collapse, or reduction, due to a second, incompatible and non-disturbing measurement.}

The state $\ket{\psi'}$ at $t_1$ depends on the interaction with the measurement apparatus $\ket{\eta}$, which prepared the system at $t_0$. Could the interaction with the apparatus $\ket{\eta}$ leave the system in the precise state $\ket{\psi'}$ which will become at $t_1$ an eigenstate of $O_1$, without a discontinuous collapse \cite{Sto08b} (fig. \ref{qm_preduction_interaction.jpg})?

\imagew{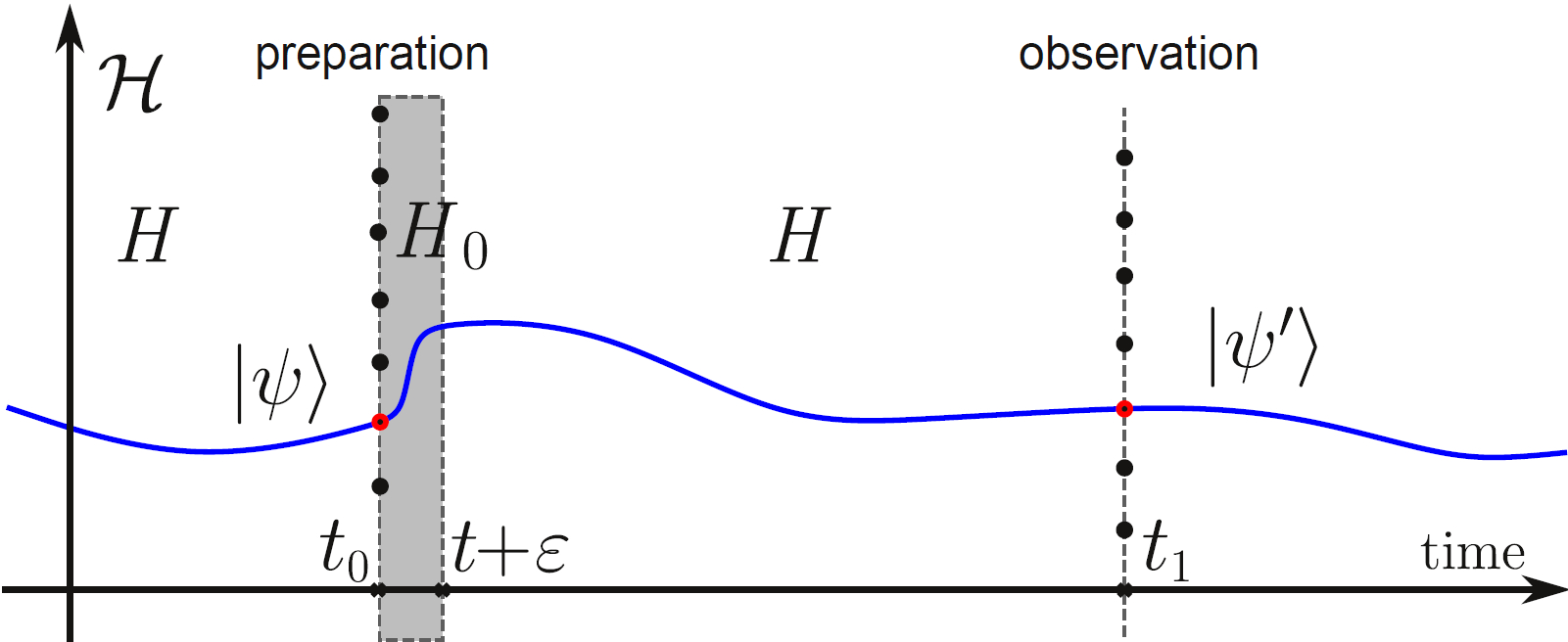}{0.65}{Could the interaction with the apparatus leave the system in the precise state $\ket{\psi'}$ which will be detected later, without a discontinuous collapse?}

Here is a possible explanation, involving only unitary evolution. The apparatus $\ket{\eta}$ is interacting with the observed system $\ket{\psi}$, and couples with it in a larger system $\ket{\psi}\ket{\eta}$. If we consider the superposition of all possible states of the observed system, it turns out that it is entangled with the apparatus (fig. \ref{qm_preduction_selection.jpg}).

\imagew{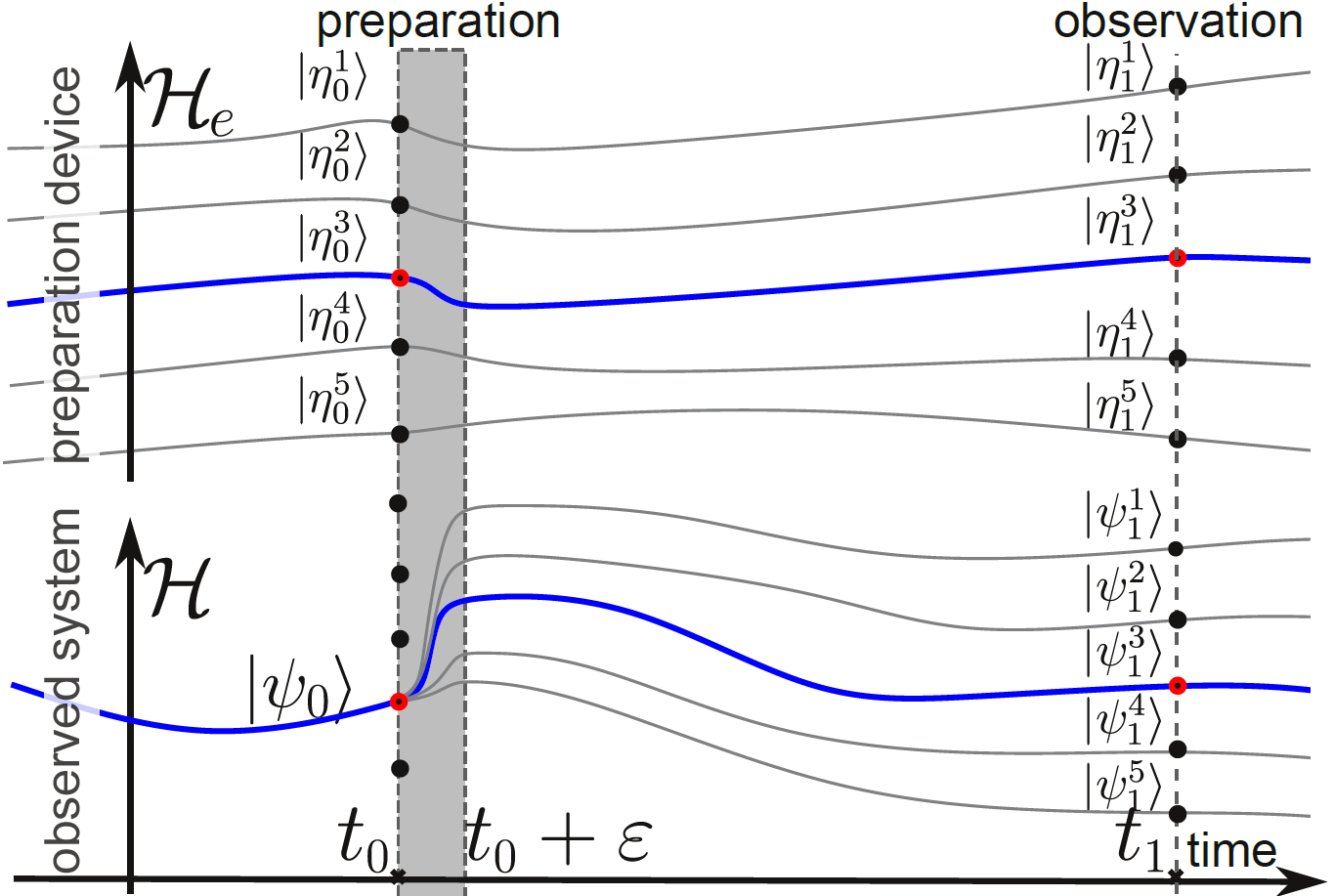}{0.65}{Each possible outcome of the measurement can be explained by a particular initial state of the observed system plus the preparation device.}

The first measurement device being a macroscopic system, its quantum state $\ket{\eta}$ \textit{is incompletely specified} at $t_0$.
The observation of $O_1$ can't tell the initial state of $\ket{\psi}$ at $t_0$, but only that the composite system $\ket{\psi}\ket{\eta}$ was at $t_0$ in a state which evolved into $\ket{\psi_1(t_1)}\ket{\eta_1(t_1)}$, where $\ket{\psi_1(t_1)}$ is an eigenstate of the observable $O_1$.

The measurement of $O_1$ refines the initial conditions of the system $\ket{\psi}$, but also those of the apparatus $\ket{\eta}$ \cite{Sto08b,Sto08f,Sto12QMa}. This is  visible also in the delayed choice experiment with the Mach-Zehnder interferometer, where the first beam splitter disturbed the photon exactly as needed to be compatible with the type of observation chosen with a delay (fig. \ref{mz_both_ways.jpg}, \ref{mz_which_way.jpg}).

The two measurements impose conditions which are incompatible if the observed system is never disturbed. But if it is disturbed, the two observations can become compatible.
From the viewpoint of time evolution, the previous interactions ``conspire'' to lead to a compatible outcome.
From the viewpoint of global consistency, the only allowed global solutions of the {\schrod} equation (for the observed system and the environment) are those which satisfy both observations, no matter how counterintuitive is this from the viewpoint of classical causality.

These examples show that even in the case of successive incompatible measurements, it is not sure that the {\qmR} process violates the {\qmU} process.

\section{Local pieces of a global puzzle}

When measured, the spin has to be $\ket{\uparrow}$ or $\ket{\downarrow}$, as in fig. \ref{sg_spin.png}. When we combine two measurements, as in fig. \ref{sg_spin_rot.jpg}, not all combinations of outcomes are valid (fig. \ref{sg_spin_rot_bad.jpg}).

\imagew{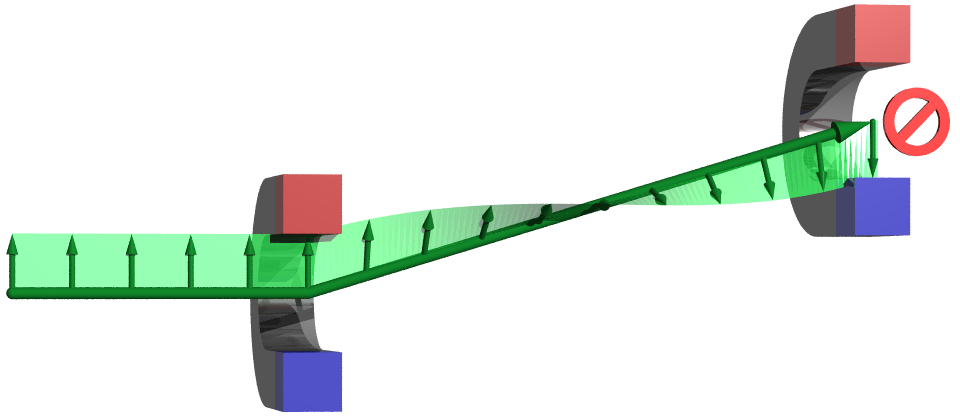}{0.65}{Two measurements along the same direction in space can't have different outcomes.}

Suppose a particle of spin zero decays into two particles of spin one-half. Spin conservation requires the resulting particles to have opposite spins. In the Einstein-Podolsky-Rosen experiment \cite{EPR35}, Alice and Bob are free to chose independently the directions along which they measure the spin (fig. \ref{puzzle_epr_rot.png}).
But if they choose the same direction, the outcomes have to be opposite (fig. \ref{puzzle_epr_same.png}).

\imagew{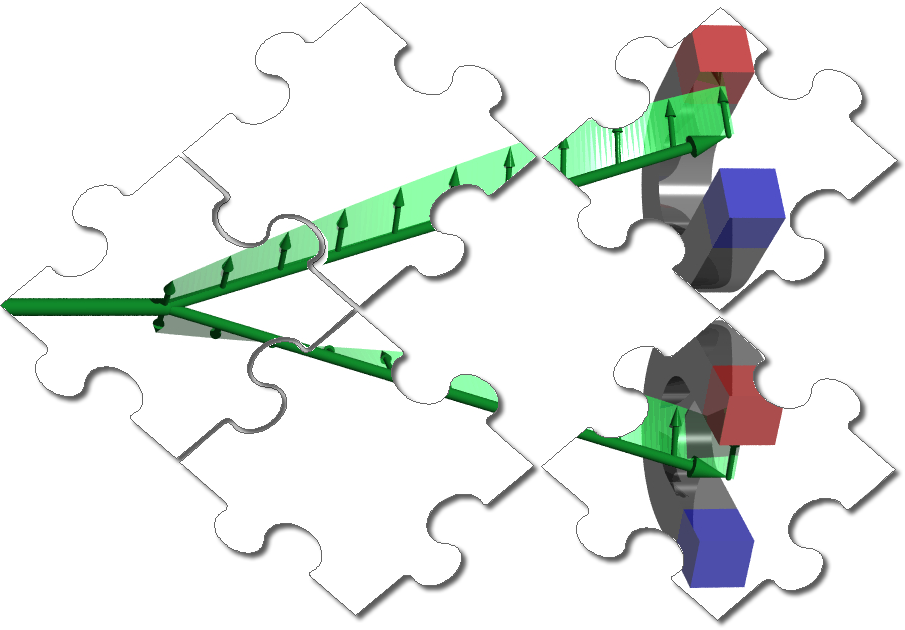}{0.55}{Alice and Bob can measure the spin along independent directions.}

\imagew{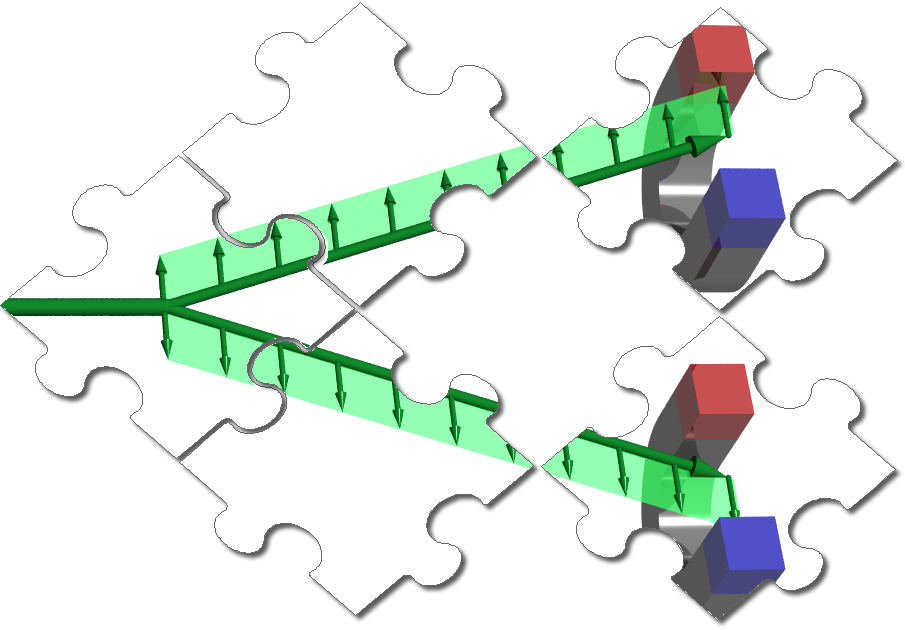}{0.55}{If Alice and Bob measure the spin along the same direction, they can't get the same value.}

The local solutions seem to glue together contiguously along a path which went back in time from Alice to the decay event, and then back to the future at Bob (or inversely).

Events like interactions (a spin 0 particle decays into two particles of spin one-half, the beam splitter splits, reflects, or transmits a photon), or measurements (the spin is up, or down, the photon is found in detector A, or B) can be seen as pieces of a puzzle. They are local solutions of {\schrod}'s equation.
Global consistency selects the possible ways to glue them to obtain larger solutions. Two pieces of the puzzle should be consistent even if they are space-like separated (fig. \ref{puzzle_epr_rot.png}), and even if one is in the past of the other (fig. \ref{puzzle_delayed.png}).

\imagew{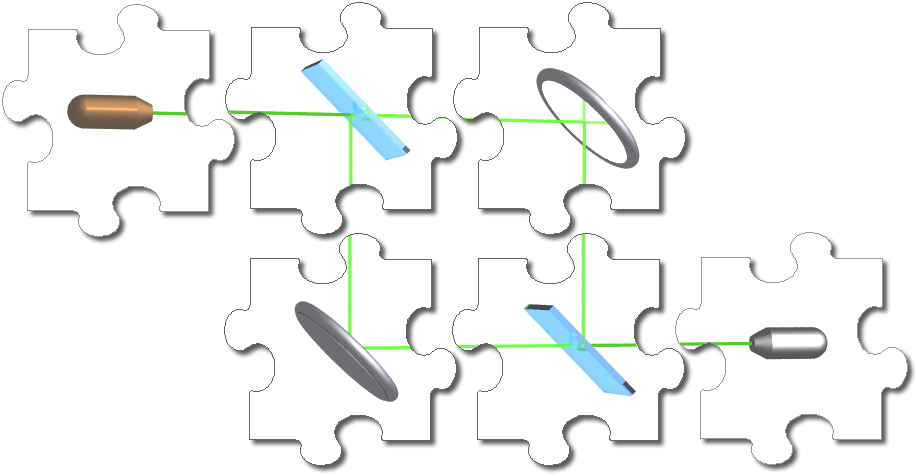}{0.7}{A puzzle piece has to be consistent even with pieces from its past or future.}

\section{Conclusions}

The proposed global perspective is far from solving the major problems of quantum mechanics, but it suggests an alternative way to think about them. It sheds a new light on the causality as normally understood, proposing instead to use notions like local and global consistency. The interactions are made again local, and the non-locality is viewed as a consequence of global consistency.

Of course, explaining the collapse only in terms of unitary evolution requires much work, in particular in obtaining the Born rule, and possibly the explanation will be less simple than admitting a discontinuous {\qmR} process.

Many connections can be made with other talks at this conference, in particular with Eliahu Cohen's and Seth Lloyd's awesome presentations. 

I cordially thank Eliahu Cohen, Hans-Thomas Elze, and Florin Mol\-do\-vea\-nu, for very helpful comments and suggestions.
I thank the organizers for this opportunity, and for the wonderful and inspiring week spent in Turin.

\end{document}